\documentclass[]{article}
\usepackage{emulateapj}
\usepackage{graphicx}

\advance \voffset by -.50cm\relax

\begin{document}

\title{The lowest-mass stellar black holes: catastrophic death of
neutron stars in gamma-ray bursts}

\author{K.\ Belczynski$^1$, R.\ O'Shaughnessy$^2$, V. Kalogera$^3$, 
        F. Rasio$^3$, R. E. Taam$^3$, T. Bulik$^4$}
\affil{$^1$ Los Alamos National Laboratory (Oppenheimer Fellow)}
\affil{$^2$ Penn State University}
\affil{$^3$ Northwestern University}
\affil{$^4$ Warsaw University}

\begin{abstract}
Mergers of double neutron stars are considered the most likely progenitors for 
short gamma-ray bursts. Indeed such a merger can produce a black hole with a 
transient accreting torus of nuclear matter (Lee \& Ramirez-Ruiz 2007, Oechslin 
\& Janka 2006), and the conversion of a fraction of the torus mass-energy to 
radiation can power a gamma-ray burst (Nakar 2006). Using available binary pulsar 
observations supported by our extensive evolutionary calculations of double 
neutron star formation, we demonstrate that the fraction of mergers that can form 
a black hole -- torus system depends very sensitively on the (largely unknown) 
maximum neutron star mass. We show that the available observations and models put 
a very stringent constraint on this maximum mass under the assumption that a 
black hole formation is required to produce a short gamma-ray burst in a double 
neutron star merger. Specifically, we find that the maximum neutron star mass 
must be within $2 - 2.5 M_\odot$. Moreover, a single unambiguous measurement of a 
neutron star mass above $2.5 M_\odot$ would exclude a black hole -- torus central 
engine model of short gamma-ray bursts in double neutron star mergers. Such an
observation would also indicate that if in fact short gamma-ray bursts are 
connected to neutron star mergers, the gamma-ray burst engine is best explained 
by the lesser known model invoking a highly magnetized massive neutron 
star (e.g., Usov 1992; Kluzniak \& Ruderman 1998; Dai et al. 2006; Metzger, 
Quataert \& Thompson 2007).
\end{abstract}

\section{Introduction}

Gamma-ray bursts (GRBs) have been separated into two classes: long-soft bursts, 
and short bursts (Nakar 2006,{Gehrels} {et~al.} 2007). 
The origin of long-soft bursts has been 
connected to the death of low-metallicity massive stars
(Piran 2005, {Gehrels} {et~al.} 2007).  
However, while observations support a binary merger origin for short bursts
(Nakar 2006,{Gehrels} {et~al.} 2007), the exact nature of the 
progenitor remains uncertain: 
they could be either double neutron stars (NS--NS) or  black hole -- neutron star
(BH--NS) binaries. The number of BH--NS binaries that both merge and produce GRBs 
is hard to estimate since (i) no such system has yet been observed; (ii) formation 
models are rather uncertain and predict very small BH--NS merger rates (likely too 
small to explain most of the short bursts); and (iii) theory suggests that the 
fraction of BH--NS mergers producing bursts depends sensitively on the black hole 
spin and spin--orbit orientation 
({Belczynski} {et~al.} 2007b), but black hole birth
spins are not well constrained observationally or theoretically. On the other hand, 
NS--NS binaries are only observed in the Milky Way, but their properties and 
numbers are also in agreement with theoretical models, and their merger rate
is sufficient to explain the present-day short burst population (Nakar 2006,
Belczynski et al. 2007a).

We have performed an extensive theoretical study of high-mass binary stars 
(potential progenitors of NS--NS systems) using \texttt{StarTrack}, a population 
synthesis code incorporating the most up-to-date and detailed input physics for 
massive stars (Belczynski et al 2008). The code employs state-of-the-art predictions
for neutron star and black hole masses based on hydrodynamic core collapse 
simulations (Fryer \& Kalogera 2001)
 and detailed stellar structure and
evolution calculations for massive stars (Timmes, Woosley, \& Weaver 1996). Our models 
predict a Galactic NS--NS merger rate in the range $\sim 10-100 {\rm Myr}^{-1}$
(Belczynski et al. 2007a), in good agreement with the empirical estimate of $\sim 3-190 
{\rm Myr}^{-1}$ ({Kim}, {Kalogera}, \&
  {Lorimer}, 2006). The spread in our predicted rates originates 
from including the most significant model uncertainties associated with the 
treatment of dynamical mass transfer episodes (common envelope phases), which are 
involved in the formation of most double compact objects (Belczynski et al. 2007a).

\section{Results}

\begin{figure}
\includegraphics[width=\columnwidth]{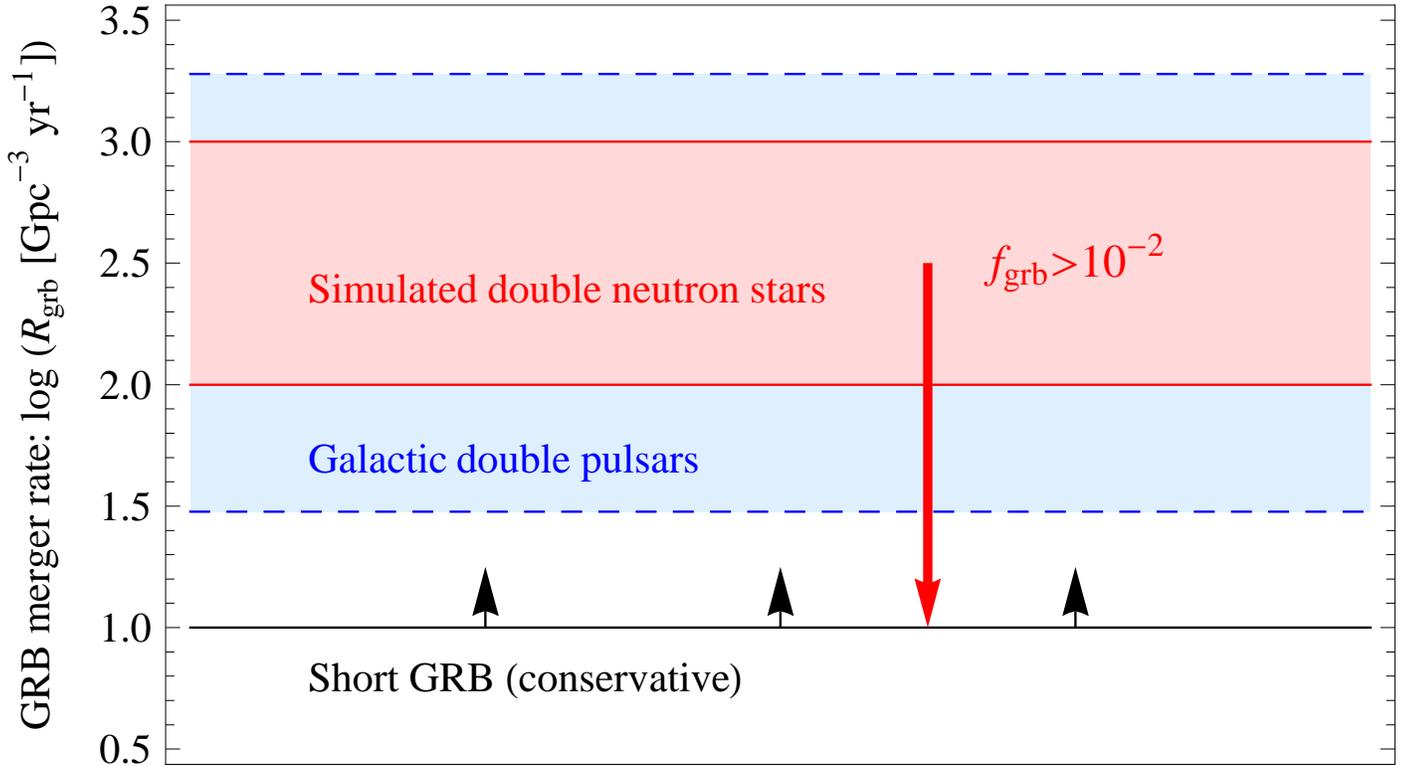}
\caption{\label{fig1} 
Comparison of the double neutron star merger rates and short GRB event rates.  
The solid black line and arrows indicate a firm lower bound on the short GRB  
event rate (Nakar 2006), based solely on the rate of detected bursts. Depending 
on the amount of beaming and the fraction of distant faint short GRBs that are 
missed, the true event rate is often estimated to be at least $10$ times larger 
(Nakar 2006). This lower limit is smaller than the double neutron star merger 
rate estimated for the Milky Way both from (i) observations of Galactic binary 
pulsars (filled blue region) and (ii) our population synthesis simulations 
(filled red region), when these two estimates are extrapolated to cosmological 
scales. Based on the maximum plausible double neutron star merger rate with the 
minimum plausible short GRB event rate, the fraction $f_{\rm grb}$ of binary mergers 
that lead to short GRBs should be greater than $10^{-2}$ if double neutron stars 
are the progenitors of short GRBs.
}
\end{figure}

In Figure \ref{fig1} we compare short GRB rates with NS--NS merger rates in the 
present-day (redshift $0$) universe. Extrapolating the NS--NS merger rates to the 
local universe by assuming a star-forming density of $10^{-2}$ Milky 
Way-equivalents per ${\rm Mpc}^3$ \footnote{Introduction of different modes of star 
formation and redistribution of progenitors over old elliptical and younger spiral 
galaxies does not increase the predicted merger rates by more than a factor of 
$\sim 3$ O'Shaughnessy et al. 2007.}, we estimate the local universe 
NS--NS merger rate to be in the range $\sim 100-1000 {\rm Gpc}^{-3} 
{\rm yr}^{-1}$. By comparison, the estimated conservative lower limit on the 
short GRB rate is $\sim 10 {\rm Gpc}^{-3}{\rm yr}^{-1}$, based on the BATSE/SWIFT 
sample (Nakar 2006). This estimate relies on very conservative assumptions: (i) 
there is no collimation and (ii) there are no bursts dimmer than we have already 
observed, thus providing a true lower limit on the rate. Therefore, even adopting 
the most optimistic predictions for the NS--NS merger rate and the most pessimistic 
bound on the local short GRB event rate, the fraction of NS--NS mergers $f_{grb}$ 
that produce GRBs must be greater than at least $10^{-2}$ to explain the majority 
of known short bursts.

In this paper we start with the assumption  that all short GRBs are connected with
NS-NS system mergers that produce a black hole. We discuss the implications of  
relaxing this stringent assumption at the end of the paper.

\begin{figure}
\includegraphics[width=\columnwidth]{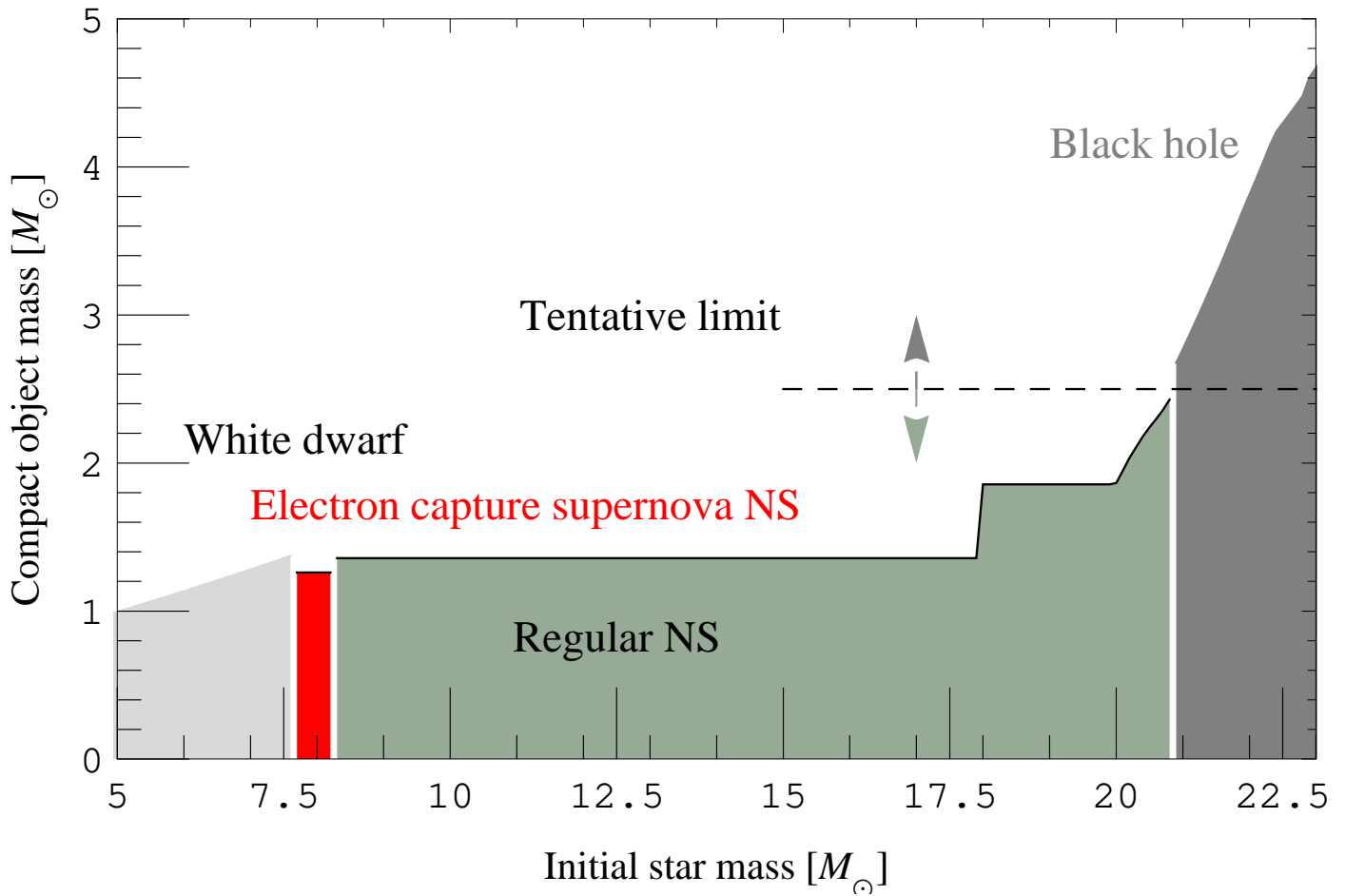}
\caption{\label{fig2}
Initial (Zero Age Main Sequence) mass to final compact object mass relation for 
single stars. This represents our current understanding of compact object 
formation. Stars below about $7.5 M_\odot$ form white dwarfs; stars in the narrow 
range around $8 M_\odot$ can potentially form very light neutron stars through 
electron capture supernovae (Podsiadlowski et al. 2004). More massive stars show a well defined 
bifurcation caused by different modes of energy transport in the stellar core: 
stars below $18 M_\odot$ form light neutron stars ($\simeq 1.35 M_\odot$), while 
stars above this mass form heavy neutron stars ($\simeq 1.8 M_\odot$). Above 
$\simeq 20 M_\odot$ stars experience partial fallback of material that can turn 
nascent neutron stars into black holes. Compact objects originating from stars 
of $\sim 20-22 M_\odot$ form either very heavy neutron stars or low-mass black 
holes depending on the unknown limiting mass between these two remnant types 
(expected to lie around $2-3 M_\odot$).
}
\end{figure}

From our models we also derive physical properties of double neutron stars, with 
individual masses of neutron stars being of particular interest. Figure 
\ref{fig2} shows the relation between progenitor (single star) mass and final 
remnant mass used in our evolutionary calculations. Mass transfer and other 
binary interactions change this simple picture, through both accretion and mass
loss, which can either increase or decrease an individual binary component mass.
However, (Belczynski et al. 2007b) argues that we do not expect significant mass 
accretion onto the components of NS--NS binaries. The population model we adopt 
for our discussion here produces NS mass distributions that appear consistent  
with the current observed NS-NS sample, at least in the extent of the mass 
ranges (Fig.\ 3). While mass transfer does influence the remnant masses (e.g., 
smearing the narrowly peaked mass distribution implied by Fig.\ 2), the 
qualitative structure is largely preserved, as one would expect from isolated 
stellar evolution combined with an initial mass function that falls steeply with 
increasing initial mass. The predicted neutron star mass distribution only 
very weakly depends on evolutionary model assumptions because the neutron
star formation mass is almost independent of the progenitor mass (Timmes et al.
1996) and mass accretion in NS-NS progenitor binaries is rather small 
(Belczynski et al. 2007a). 

Depending on the masses in the progenitor binary and the highly uncertain 
nuclear equation of state, the final remnant of a NS--NS merger may or may not 
collapse to a black hole. We estimate the final gravitational mass of the compact remnant as
\begin{equation}
\label{eq:FinalRemnantMass}
M_{\rm rem} = 0.9 (M_{\rm ns,1}+M_{\rm ns,2} - 0.1 M_\odot),
\end{equation}
where the initial neutron star masses are denoted by $M_{\rm ns,1},\ 
M_{\rm ns,2}$ and we have assumed that the torus mass is sufficiently large to 
power a GRB (i.e., $\simeq 0.1 M_\odot$)
({Setiawan}, {Ruffert}, \&
  {Janka} 2004, Lee \& Ramirez-Ruiz 2007)
and that 10\% of rest mass is lost in neutrinos (Lattimer \& Yahli 1989; Timmes et
al. 1996). Higher rest mass loss (e.g., Metzger, Thompson \& Quataert 2007) would only
strengthen our subsequent conclusions. Because stars more massive than $18 
M_\odot$ (progenitors of massive neutron stars with $M_{\rm ns} \simeq 1.8 M_\odot$) 
are much rarer than those forming lighter neutron stars ($M_{\rm ns} \simeq 1.35 
M_\odot$) we \emph{a priori} expect that most remnants from NS--NS mergers will 
have rather low mass $M_{\rm rem} \simeq 2.3 M_\odot$ (see eq.~1 for two $1.35 M_\odot$
neutron stars).

Neither observations nor nuclear theory have yet pinned down the maximum 
neutron star mass $M_{\rm ns,max}$ above which a black hole must form. Thus, the 
fraction of binary neutron stars which produce black holes and are able to 
power short GRBs is set by the fraction of mergers such that
\begin{equation}
M_{\rm rem} \geq M_{\rm ns,max}.
\end{equation}
We therefore calculate the fraction of our simulated NS--NS mergers that 
lead to black hole formation and a short GRB as a function of $M_{\rm ns,max}$; 
see Fig.~4. Observations of the highest mass neutron stars ($\lesssim 2 
M_\odot$;
( {Barziv} {et~al.} 2001, {Ransom} {et~al.} 2005)
 and lowest mass 
black holes ($\gtrsim 3 M_\odot$; 
(Orosz 2003, Casares 2006) 
only weakly 
constrain this parameter. Remnant masses from NS--NS mergers ($M_{\rm rem}$) 
obtained both from our simulations and from observations all fall very 
close to the range $2.2-2.5 M_\odot$.

Comparing Figs 1 and 4 we immediately deduce that, since the fraction $f_{\rm grb}$
of NS--NS mergers that produce short GRBs must be greater than $10^{-2}$ 
(Fig.~1), the neutron star maximum mass $M_{\rm ns,max}$ must be less than $2.5 
M_\odot$ (Fig.~4). Because we lack a robust lower bound on the mass of the 
residual torus surrounding the black hole, we have adopted a conservative 
upper limit on $M_{\rm ns,max}$ obtained by assuming a negligible torus mass (i.e., 
replace $0.1 M_\odot$ with $0$ in eq.~1).

\begin{figure}
\includegraphics[width=\columnwidth]{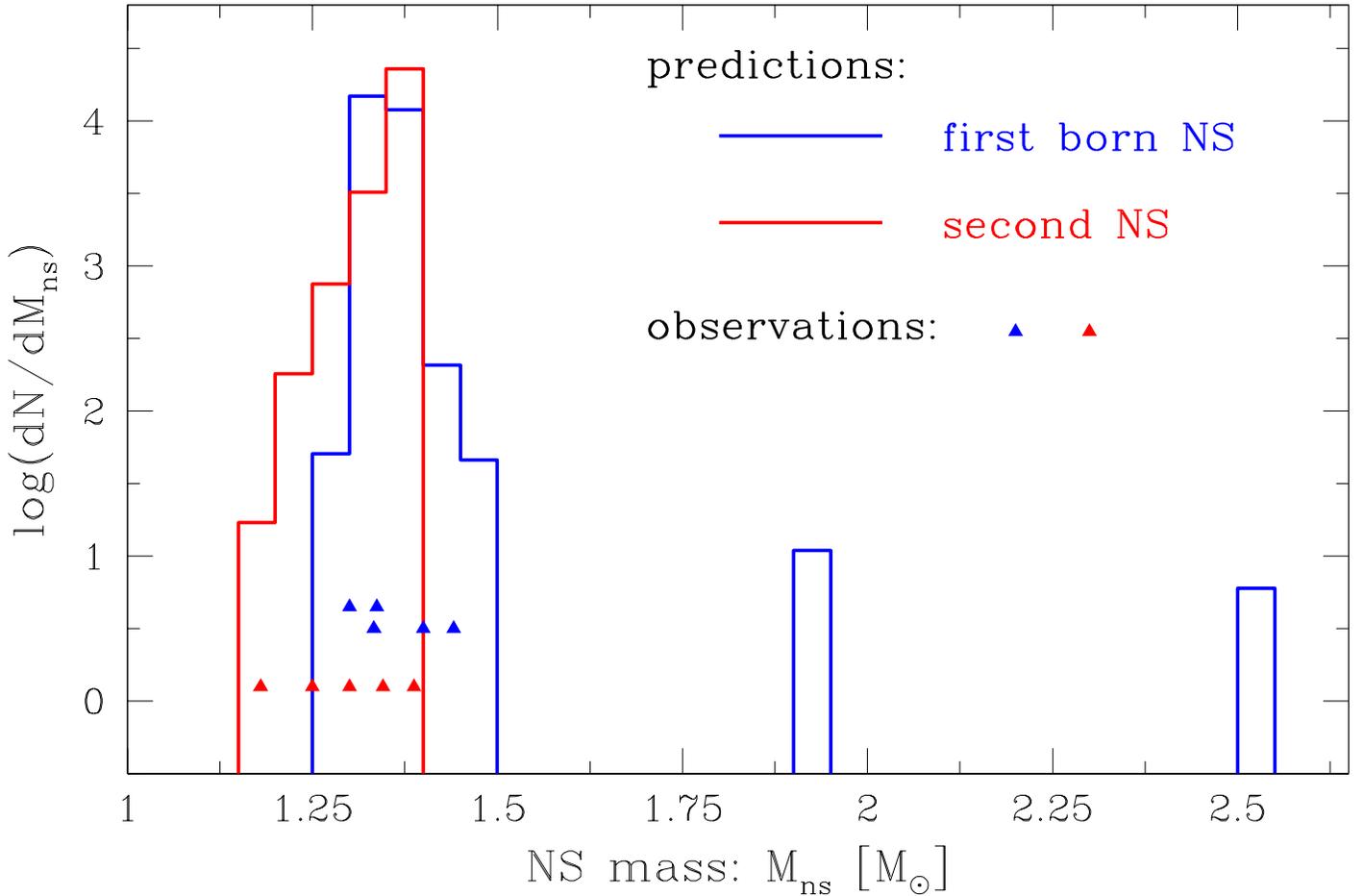}
\caption{\label{fig3}
Predicted mass distribution for neutron stars in merging double neutron star 
binaries. First born neutron stars are slightly heavier as they can accrete 
some matter from their unevolved binary companions. Population synthesis 
models (red and blue lines) are shown along with measured neutron star masses 
for the known double neutron star binaries. Although more observations are 
needed to constrain the shape of this distribution, the mass ranges of 
observed and predicted systems are in agreement. We use direct mass estimates 
for B1913$+$16, B1534$+$12, J0737$-$3039 and J1756$-$2251 (O'Shaughnessy et
al. 2008), 
while for J1906$+$0746 we assume that both neutron stars have masses of $1.3 
M_\odot$ (total system mass is $2.6 M_\odot$; Lorimer et al. 2006). The 
few compact objects found in our simulations with masses as high as $\simeq 
2.5 M_\odot$ may well be low-mass black holes (see also Fig.~\ref{fig1}).
}
\end{figure}

This result has been obtained with the assumption that all ($k=1.0$) short GRBs are 
connected with NS-NS mergers. It is however possible that only a fraction of short 
GRBs is produced in NS-NS mergers. How does our result depend on the fraction $k$ of 
short GRBs that are connected with NS-NS mergers?
The lower limit on $f_{\rm grb}$ is then decreased by $k$, see Figure~\ref{fig1}.
If $k \gtrsim 0.1$, then the limit lower limit becomes $f_{\rm grb}>10^{-3}$, and 
as is clearly seen from Figure~\ref{fig3}, the upper limit on the maximum 
mass of a neutron star remains unchanged. 
For the values of $k \lesssim 0.01$ the NS-NS mergers are not important for
overall short GRB population, as the mergers would consist of only $\lesssim
1\%$ of the short GRBs. Therefore, in this case short GRBs do not provide
information about the merger product.

\section{Discussion}

Our proposed limit on the maximum neutron star mass is still above the maximum 
masses allowed by almost all proposed models for the nuclear equation of state 
({Lattimer} \& {Prakash} 2007). However, the proposed limit would remain unchanged 
even if a dramatic improvement in short GRB surveys led to a significantly larger 
lower bound on the local short GRB rate:  because of the sharp decrease in 
$f_{\rm grb}$ with $M_{\rm ns,max}$ shown in Fig.~4.
If, however, electromagnetic observations could constrain the 
least luminous short GRBs and thus provide an \emph{upper} bound on the short 
GRB rate, with gravitational wave observations at the same time accurately 
determining the NS--NS merger rate, then $f_{\rm grb}$ could also be constrained from 
\emph{above}. If only a fraction of NS--NS mergers produce short bursts, 
because $f_{\rm grb}$ depends so sensitively on $M_{\rm ns,max}$, the combination of upper 
and lower limits would constrain the maximum neutron star mass extremely 
tightly, even if the assumptions going into eq.~1 are relaxed. 

While our limit on the effective maximum neutron star mass is entirely 
empirical, detailed merger models including realistic relativistic dynamics, 
neutrino transport, magnetic fields, and potentially even energy extraction
from the final black hole remain under intense investigation (Janka \& Ruffert 
1996; Oechslin \& Janka 2006). Many merger remnants are 
expected to be (temporarily) rotationally supported against collapse 
(Morrison, Baumgarte \& Shapiro 2004), with a ``hypermassive'' remnant neutron star 
eventually spinning down and collapsing to a black hole 
(Faber et al. 2006; Duez et al. 2007; Shibata \& Taniguchi 2006).
Our model only relies upon the current consensus on double neutron star 
mergers, as summarized by Oechslin \& Janka (2006): 
sufficiently massive binary mergers produce a black hole and only mergers 
that produce a black hole extract enough energy to power short GRBs.

Known Galactic black holes extend in mass up to $10-15 M_\odot$ (Casares 2006),
while two recently discovered black hole candidates in other galaxies 
(Orosz et al. 2007, Prestwich et al. 2007)  have even higher masses of 
$\simeq 16$ and $\gtrsim 24 M_\odot$. Clearly black holes can form with rather 
high masses in different types of environments. The lower mass limit is not 
well constrained observationally, as the highest-mass neutron stars barely 
reach $2 M_\odot$, while the lowest-mass black holes are above $3 M_\odot$. 
In order to explain the observed short GRBs with NS--NS mergers, under the 
assumption of black hole -- torus central engine model, we have shown that 
the maximum neutron star mass must be lower than $2.5 M_\odot$. 
However, pulsar surveys (Ransom et al. 2005) have discovered increasingly
more massive neutron stars. So far in our analysis we 
have included only NS--NS systems formed in the field.  There is one known 
relativistic double neutron star system in the Galactic globular cluster M15, 
that has probably formed through dynamical interactions.  This binary consists 
of two low-mass neutron stars ($1.36$ and $1.35 M_\odot$; Jacoby et al. 2006)
very similar to those in the Galactic field, so the results of our analysis 
are not changed by this isolated observation. Moreover, it was estimated that
no more than $10-30\%$ short GRBs can originate from mergers of double neutron 
stars formed in globular clusters ({Grindlay}, {Portegies Zwart}, \& {McMillan} 
2006).

If any observation can be made that establishes unambiguously a pulsar mass 
(either in the field or a globular cluster) over $2.5 M_\odot$, this would 
exclude a black hole -- torus short GRB central engine model for double neutron 
star mergers. We note that a tentative mass measurement for a pulsar of 
$2.74 \pm 0.21 M_\odot$ was recently reported by (Freire et al. 2007). If this 
measurement is confirmed, the double neutron star mergers may still be possible 
progenitors for short GRBs. However, the central engine model will need to be 
reexamined.  In particular, it was proposed that a merger of two neutron stars 
may lead to the formation of a magnetar; a rapidly rotating highly 
magnetized and high mass neutron star (with or without a torus) that can lead 
to a short GRB (e.g., Usov 1992; Kluzniak \& Ruderman 1998; Dai et al. 2006; 
Metzger et al. 2007).  Gravitational wave observatories (LIGO, VIRGO) may 
provide the direct evidence that NS-NS merger can produce a short GRB if there 
is a coincidence of the burst and the inspiral gravitational wave signal. 
It may also be possible to distinguish a merger product (NS versus BH) 
from the shape of the merger and ringdown signal or from radio pulses if a 
magnetar was formed in a nearby gamma-ray burst event.  

\begin{acknowledgements}
We would like to thank Neil Gehrels, Duncan Lorimer, Scott Ransom, Ben Owen,
Jerome Orosz, Chris Stanek, John Beacom, Paulo Freire, Chunglee Kim, Todd
Thompson and an anonymous referee for very useful discussions. 
\end{acknowledgements}

\begin{figure}
\includegraphics[width=1.05\columnwidth]{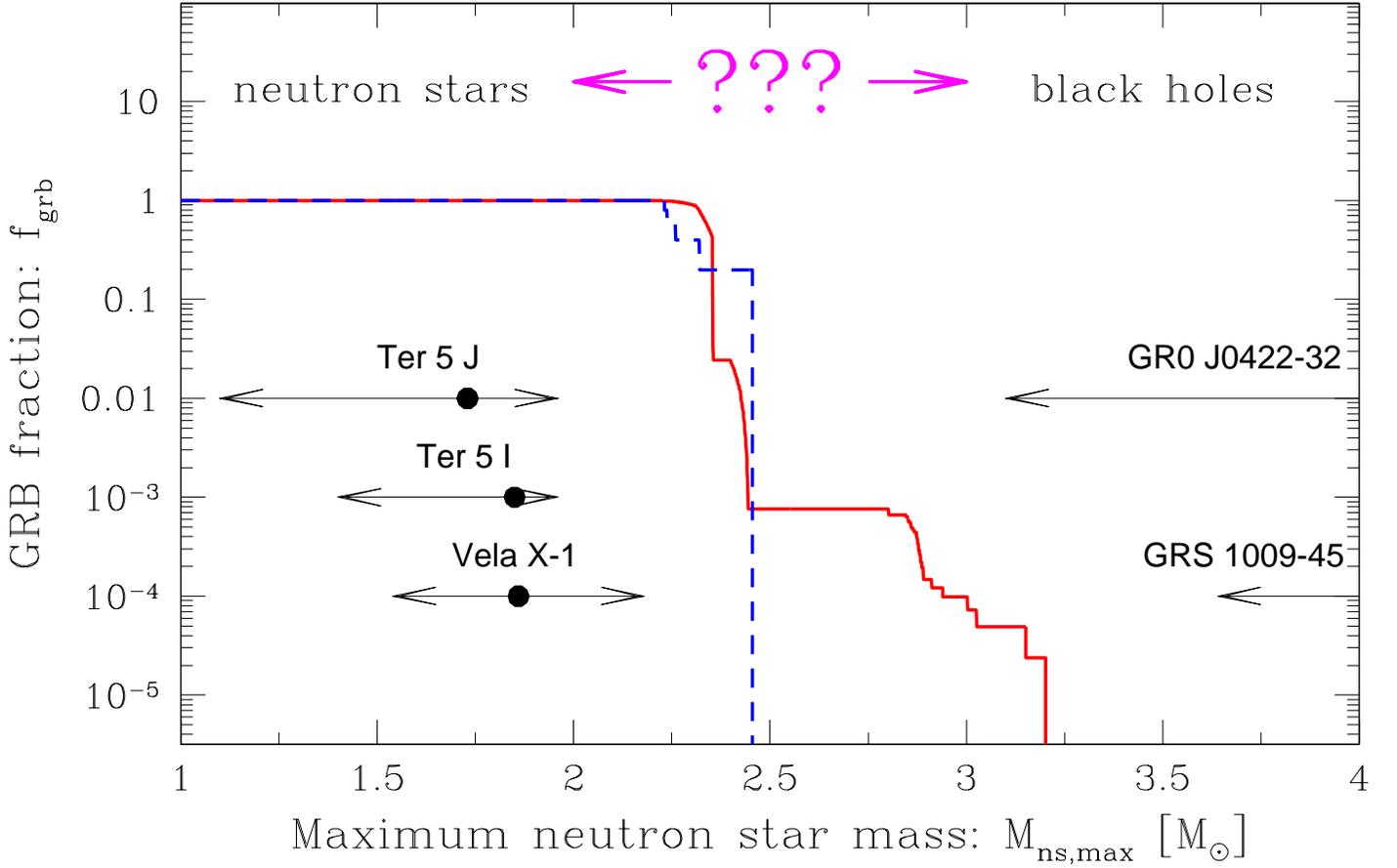}
\caption{\label{fig4}
Gamma Ray Burst production efficiency as a function of the maximum neutron 
star mass in the framework of the double neutron star model for GRBs 
involving the formation of a black hole. 
The mass of the merger product is plotted 
here as the blue (observations) and red (theory) lines. There is a sharp drop 
in number of NS--NS systems that can form a merger with mass over $2.5 M_\odot$. 
For example, only 1 in $10^3$ NS--NS mergers can form a remnant with a mass 
of $2.5 M_\odot$ or higher. Therefore, if short GRBs are connected to NS--NS 
mergers, the maximum neutron star mass is required to be $M_{\rm ns,max} < 2.5 M_\odot$.
For comparison we show observed masses of the lowest-mass black holes (GRO 
J0422$-$32, GRS 1009$-$45; Casares 2006) and highest-mass neutron stars (Vela X-1, 
Terzan~5I, and Terzan~5J; Barziv et al. 2001, Ransom et al. 2005).
These observations, along with our findings, constrain the maximum mass of a neutron 
star to lie in the narrow range of $2-2.5 M_\odot$.
}
\end{figure}

\end{document}